\documentclass[12pt]{article}%
\usepackage{a4}
\usepackage{amsfonts,amssymb,amsmath}
\usepackage{graphicx}
\usepackage{amsmath}
\usepackage{amsfonts}
\usepackage{amssymb}%
\begin{document}

\title{Position-dependent noncommutativity in quantum mechanics}
\author{M.Gomes\thanks{e-mail: mgomes@fma.if.usp.br}, V.G. Kupriyanov\thanks{e-mail:
vladislav.kupriyanov@gmail.com}\\
\\Instituto de F\'{\i}sica, Universidade de S\~{a}o Paulo, Brazil}
\date{\today                                        }
\maketitle

\begin{abstract}
The model of the position-dependent noncommutativety in quantum mechanics is
proposed. We start with a given commutation relations between the operators of
coordinates $\left[  \hat{x}^{i},\hat{x}^{j}\right]  =\omega^{ij}\left(
\hat{x}\right)  $, and construct the complete algebra of commutation
relations, including the operators of momenta. The constructed algebra is a
deformation of a standard Heisenberg algebra and obey the Jacobi identity. The
key point of our construction is a proposed first-order Lagrangian, which
after quantization reproduces the desired commutation relations. Also we study
the possibility to localize the noncommutativety.

\end{abstract}

\section{Introduction}

Recently quantum field theory on noncommutative spaces has been studied
extensively, see e.g. \cite{NCreviews} and references therein. General quantum
mechanical arguments indicate that it is not possible to measure a classical
background space-time at the Planck scale, due to the effects of the
gravitational backreaction \cite{Doplicher}. This has led to the belief that
the classical differentiable manifold structure of space-time at the Planck
scale should be replaced by some sort of noncommutative structure. The
simplest approximation is a flat noncommutative space-time, which can be
realized by the coordinate operators $\hat{x}^{\mu}$ satisfying $\left[
\hat{x}^{\mu},\hat{x}^{\nu}\right]  =i\hbar\theta^{\mu\nu}\,,$ where
$\theta^{\mu\nu\,}$ is the noncommutativity parameter. However, the
restriction to flat space-time is not natural and one must discuss more
general curved noncommutative space-time, when the commutator of coordinates
depends on these coordinates. The generalized noncommutative spaces arise e.g.
in the context of string theory because of the presence of background
antisymmetric magnetic $B$-field.

The construction of a consistent quantum field theory and gravity on a curved
noncommutative space is one of the main open challenges in modern theoretical
physics. However, to do it is not so easy because of the conceptual and
technical problems. To begin with let us study quantum mechanics QM with
position-dependent noncommutativity.

Usually, noncommutative QM \cite{NCQM} deals with the following commutation
relations:
\begin{align}
\left[  \hat{x}^{i},\hat{x}^{j}\right]   &  =i\hbar\theta^{ij},\ \label{1}\\
\ \left[  \hat{x}^{i},\hat{p}_{j}\right]   &  =i\hbar\delta_{j}^{i}%
,\ \ \label{1a}\\
\left[  \hat{p}_{i},\hat{p}_{j}\right]   &  =0, \label{1b}%
\end{align}
where $\theta^{ij}$ is some constant antisymmetric matrix. However, it is not
always reasonable to assume that the noncommutativity extends to the whole
space, leaving the parameter of noncommutativity $\theta^{ij}$ to be constant.
One can consider more general situation of position-dependent or even local
noncommutativity, when noncommutativity exists only in some restricted area of
the space, like, e.g., in the two-dimensional case,
\begin{equation}
\lbrack\hat{x},\hat{y}]=\frac{i\hbar\theta}{1+\theta\alpha\left(  \hat{x}%
^{2}+\hat{y}^{2}\right)  }~. \label{2}%
\end{equation}
The constant $\alpha$ is a parameter which measure the degree of locality, if
$\alpha=0$ the noncommutativity is global (\ref{1}-\ref{1b}), if $\alpha\neq0$
the noncommutativity is local. Other examples of position-dependent
noncommutativity are Lie-algebraic $\left[  \hat{x}^{i},\hat{x}^{j}\right]
=i\hbar f_{k}^{ij}\hat{x}^{k}$ and, in particular the kappa-Poincare
noncommutativity \cite{Lukierski}, and the quadratic noncommutative algebra
$\left[  \hat{x}^{i},\hat{x}^{j}\right]  =i\hbar R_{kl}^{ij}\hat{x}^{k}\hat
{x}^{l}$ which appears in the context of quantum groups \cite{Sklyanin},
\cite{Szabo}.

The aim of this work is to construct consistent quantum mechanics with a given
position-dependent noncommutativity,%
\begin{equation}
\left[  \hat{x}^{i},\hat{x}^{j}\right]  =i\hbar\omega^{ij}\left(  \hat
{x}\right)  ~, \label{3}%
\end{equation}
i.e., to construct the complete algebra of commutation relations, including
momenta, which obey the Jacobi identity.

\section{Jacobi identity and position-dependent noncommutativity}

Note that in the presence of the position-dependent noncommutativity
(\ref{3}), the other commutators $\left[  \hat{x}^{i},\hat{p}_{j}\right]  $
and $\left[  \hat{p}_{i},\hat{p}_{j}\right]  $ should be changed as well in
order to satisfy the Jacobi identity. For example, consider the identity%
\begin{equation}
\left[  \hat{p}_{k},\left[  \hat{x}^{i},\hat{x}^{j}\right]  \right]  +\left[
\hat{x}^{j},\left[  \hat{p}_{k},\hat{x}^{i}\right]  \right]  +\left[  \hat
{x}^{i},\left[  \hat{x}^{j},\hat{p}_{k}\right]  \right]  \equiv0~, \label{4}%
\end{equation}
where coordinates obey (\ref{3}) and momenta still obey (\ref{1a}),
(\ref{1b}). Then from (\ref{4}) one has:
\[
\left[  \hat{p}_{k},\omega^{ij}\left(  \hat{x}\right)  \right]  +\left[
\hat{x}^{j},\delta_{k}^{i}\right]  +\left[  \hat{x}^{i},\delta_{k}^{j}\right]
\equiv0~,
\]
or
\begin{equation}
\left[  \hat{p}_{k},\omega^{ij}\left(  \hat{x}\right)  \right]  \equiv0~.
\label{5}%
\end{equation}
If we suppose now that
\begin{equation}
\omega^{ij}\left(  \hat{x}\right)  =f_{l}^{ij}\hat{x}^{l}~, \label{6}%
\end{equation}
then from (\ref{1a}) and (\ref{5}) it follows that%
\begin{equation}
\left[  \hat{p}_{k},f_{l}^{ij}\hat{x}^{l}\right]  =-i\hbar f_{l}^{ij}%
\delta_{k}^{l}=-i\hbar f_{k}^{ij}\equiv0~. \label{7}%
\end{equation}
Thus, because of the Jacobi identity, the NCQM commutation relations
(\ref{1}-\ref{1b}) are valid only for a position independent parameter
$\theta^{ij}$. Otherwise, we should change (\ref{1a}) and (\ref{1b}) as well
in order to satisfy the Jacobi identity including coordinates and momenta. And
the question is how to do it?

\section{The model of position-dependent noncommutativity}

To answer the question posed at the end of the previous section, let us
consider the classical model described by the first-order Lagrangian%
\begin{equation}
L=p_{i}\dot{x}^{i}-H\left(  p,x\right)  +\left(  p_{i}+B_{i}\left(
x,\alpha\right)  \right)  \theta^{ij}\left(  \dot{p}_{j}+\dot{B}_{j}\left(
x,\alpha\right)  \right)  /2~, \label{8}%
\end{equation}
where the functions $B_{i}$ depend on the parameter $\alpha$, such that
$B_{i}\rightarrow0$ if $\alpha\rightarrow0$, and $H\left(  p,x\right)  $ is a
given function which we will call Hamiltonian. This Lagrangian is, in fact, a
generalization of a first-order model \cite{Kup} which reproduce after
quantization the NCQM commutation relations (\ref{1})-(\ref{1b}). Note that
first-order Lagrangians also have been used in the context of chiral bosons
\cite{CBoson}. For simplicity we consider just a two dimensional case,
$i=1,2,$ $x^{i}=(x,y),\ p_{i}=\left(  p_{x},p_{y}\right)  ,\ B_{i}=\left(
B_{x},B_{y}\right)  $ and$\ $%
\begin{equation}
\theta^{ij}=\theta\varepsilon^{ij}, \label{9}%
\end{equation}
where $\theta$ is a real number which, as we will see, controls the
noncommutativity, and $\varepsilon^{12}=1$. In the limit of $\theta
\rightarrow0$ the action (\ref{8}) transforms into the usual Hamiltonian
action of classical mechanics.

The Hamiltonization and canonical quantization of theories with first-order
Lagrangians were considered in \cite{GK}, see also \cite{FJ}. Following the
general lines of \cite{GK}, we construct the Hamiltonian formulation of
(\ref{8}). Let us first rewrite (\ref{8}) as%
\begin{equation}
L=p_{i}\dot{x}^{i}+\frac{\theta}{2}p_{i}\varepsilon^{ij}\dot{p}_{j}+\theta
B_{i}\varepsilon^{ij}\dot{p}_{j}+\frac{\theta}{2}B_{j}\varepsilon^{jk}%
\partial_{i}B_{k}\dot{x}^{i}-H\left(  p,x\right)  ~.\label{8a}%
\end{equation}
We adopt the notation of \cite{GK}, $\xi^{\mu}=\left(  x,y,p_{x},p_{y}\right)
,\ J_{\mu}=\left(  J_{i},J_{i+2}\right)  $, where%
\begin{equation}
J_{i}=p_{i}+\frac{\theta}{2}B_{j}\varepsilon^{jk}\partial_{i}B_{k}%
,\ \ J_{i+2}=-\frac{\theta}{2}\varepsilon^{ij}\left(  p_{j}+2B_{j}\right)
~.\nonumber
\end{equation}
In this notation (\ref{8a}) has the form%
\begin{equation}
L=J_{\mu}\dot{\xi}^{\mu}-H\left(  \xi\right)  ~.\label{8b}%
\end{equation}
The Hamiltonization of the first-order Lagrangian (\ref{8b}) leads to the
Hamiltonian theory with second-class constraints
\begin{equation}
\Phi_{\mu}\left(  \xi,\pi\right)  =\pi_{\mu}-J_{\mu}(\xi)=0~,\label{constr}%
\end{equation}
where $\pi_{\mu}$ are the momenta conjugated to $\xi_{\mu}$. The constraint
bracket is
\[
\{\Phi_{\mu},\Phi_{\nu}\}=\ \Omega_{\mu\nu}=\partial_{\mu}J_{\nu}%
-\partial_{\nu}J_{\mu}~.
\]
For the canonical variables $\xi_{\mu}$ the Dirac brackets are%
\[
\left\{  \xi^{\mu},\xi^{\nu}\right\}  _{D}=\omega_{0}^{\mu\nu},\ \ \omega
_{0}^{\mu\nu}=\Omega_{\mu\nu}^{-1}~.
\]
The explicit form is:%
\begin{align}
&  \left\{  x^{i},x^{j}\right\}  _{D}=\theta d\varepsilon^{ij},\label{10}\\
&  \left\{  x^{i},p_{j}\right\}  _{D}=d\left(  \delta_{j}^{i}-\theta
\varepsilon^{ik}\partial_{k}B_{j}\right)  ,\nonumber\\
&  \left\{  p_{i},p_{j}\right\}  _{D}=\theta\left(  \partial_{2}B_{2}%
\partial_{1}B_{1}-\partial_{1}B_{2}\partial_{2}B_{1}\right)  d\varepsilon
_{ij},\nonumber
\end{align}
where%
\begin{equation}
d=\frac{1}{1+\theta\left(  \partial_{1}B_{2}-\partial_{2}B_{1}\right)
}.\label{11}%
\end{equation}
It is easy to see that in the commutative limit, $\theta\rightarrow0$, the
constructed Dirac brackets (\ref{10}) transform into the canonical Poisson
brackets $\left\{  x^{i},x^{j}\right\}  =\left\{  p_{i},p_{j}\right\}
=0,\ \left\{  x^{i},p_{j}\right\}  =\delta_{j}^{i}$, and in the limit
$\alpha\rightarrow0$ ($B_{i}\rightarrow0$), (\ref{10}) transform into%
\[
\left\{  x^{i},x^{j}\right\}  _{D}=\theta\varepsilon^{ij},\ \ \left\{
x^{i},p_{j}\right\}  _{D}=\delta_{j}^{i},\ \ \left\{  p_{i},p_{j}\right\}
_{D}=0,
\]
which will reproduce after quantization NCQM commutation relations
(\ref{1})-(\ref{1b}). So, in the general case, the vector field $B_{i}$
introduced in order to generalize the previously known model \cite{Kup}, can
be interpreted as the correction to the simplectic potential which measure the
curvature of the phase space due to noncommutativity.

At this point we may ask if it is possible to generalize the above
construction to the case of second order models, i.e., models whose
Lagrangians are quadratic in the velocities. To investigate this possibility
we consider the model introduced by Lukierski et al \cite{Lukierski2}:%
\begin{equation}
L_{LSZ}=\frac{\dot{x}_{i}^{2}}{2}+\frac{\theta}{2}\varepsilon_{ij}\dot{x}%
_{i}\ddot{x}_{j}. \label{l1}%
\end{equation}
Introducing Lagrangian multipliers $p_{i}$ and new variables $y_{i}$, one
rewrites (\ref{l1}) in an equivalent form:%
\begin{equation}
L^{\left(  0\right)  }=p_{i}\left(  \dot{x}_{i}-y_{i}\right)  +\frac{y_{i}%
^{2}}{2}+\frac{\theta}{2}\varepsilon_{ij}y_{i}\dot{y}_{j}. \label{l2}%
\end{equation}
Next, by using the Horvathy-Plyushchay variables \cite{HP}
\begin{equation}
X_{i}=x_{i}+\theta\varepsilon_{ij}y_{j}-\theta\varepsilon_{ij}p_{j}%
,\ \ Q_{i}=\theta\left(  y_{i}-p_{i}\right)  , \label{l3}%
\end{equation}
we represent (\ref{l2}) as%
\begin{equation}
L^{\left(  0\right)  }=L_{ext}^{\left(  0\right)  }+L_{int}^{\left(  0\right)
}, \label{l4}%
\end{equation}
where%
\begin{align*}
L_{ext}^{\left(  0\right)  }  &  =p_{i}\dot{X}_{i}+\frac{\theta}{2}%
\varepsilon_{ij}p_{i}\dot{p}_{j}-\frac{1}{2}p_{i}^{2},\\
L_{int}^{\left(  0\right)  }  &  =\frac{1}{2\theta}\varepsilon_{ij}Q_{i}%
\dot{Q}_{j}+\frac{1}{2\theta^{2}}Q_{i}^{2}.
\end{align*}
We see that Lagrangian (\ref{l4}) separates into two disconnected parts
describing the \textquotedblleft external\textquotedblright\ and
\textquotedblleft internal\textquotedblright\ degrees of freedom. The
Lagrangian $L_{ext}^{\left(  0\right)  }$ is exactly a first-order model
\cite{Kup} for which we construct the generalization (\ref{8}). Note that if
now to put in (\ref{l4}) instead $L_{ext}^{\left(  0\right)  }$ the
generalized Lagrangian (\ref{8}) and then to make an inverse transformation to
(\ref{l3}) (to turn back from the Horvathy-Plyushchay variables to the
original ones) we will come to a Lagrangian involving time derivatives of
variables $p_{i}$. So, $p_{i}$ are not Lagrangian multipliers any more and
cannot be eliminated from consideration in order to go back to the higher
order model (\ref{l1}). Therefore, the generalization to the case of an
arbitrary fields $B_{i}$ is possible only in the first-order model \cite{Kup}.

\section{Quantization}

After canonical quantization, the Dirac brackets (\ref{10}) will determine the
commutation relations between the operators of the coordinates and momenta
$\hat{\xi}^{\mu}=\left(  \hat{x},\hat{y},\hat{p}_{x},\hat{p}_{y}\right)  $:
\begin{equation}
\left[  \hat{\xi}^{\mu},\hat{\xi}^{\nu}\right]  =i\hbar\omega^{\mu\nu}\left(
\hat{x},\hat{y}\right)  ,\label{11q}%
\end{equation}
and quantum Hamiltonian $\hat{H}$ is constructed according to the classical
function $H\left(  p,x\right)  ,$where some ordering must be chosen in order
to construct the operators $\omega^{\mu\nu}\left(  \hat{x},\hat{y}\right)  $
and $\hat{H}$. The most natural choice is the symmetric Weyl ordering
prescription, where to each function $f\left(  \xi\right)  $ on the phase
space is associated a symmetrically ordered operator function $\hat{f}\left(
\hat{\xi}\right)  $ according to the rule%
\begin{equation}
\hat{f}\left(  \hat{\xi}\right)  =\int\frac{d^{4}k}{\left(  2\pi\hbar\right)
^{4}}\tilde{f}\left(  k\right)  e^{-\frac{i}{\hbar}k_{\mu}\hat{x}^{\mu}%
},\label{12q}%
\end{equation}
with $\tilde{f}(k)$ is a Fourier transform of $f$. In particular, the function
$d\left(  x,y\right)  $ will determine the position-dependent
noncommutativety, $[\hat{x},\hat{y}]=i\hbar\theta d\left(  \hat{x},\hat
{y}\right)  $.

In \cite{KV} it was shown that the Jacobi identity for the operator algebra
(\ref{11q}) is equivalent to the following condition%
\begin{equation}
\left(  \xi^{\mu}\star\omega^{\nu\lambda}-\omega^{\nu\lambda}\star\xi^{\mu
}\right)  +\mbox{cycl}(\mu\nu\lambda)=0~,\label{i1}%
\end{equation}
where%
\begin{equation}
f\star g=\sum_{k=0}^{\infty}\hbar^{k}f\star_{k}g=f\cdot g+\frac{i\hbar}%
{2}\omega^{\mu\nu}\partial_{\mu}f\partial_{\nu}g+...\label{i2}%
\end{equation}
is a star product associated with the noncommutative algebra (\ref{11q}) and
$\omega^{\nu\lambda}=\omega_{0}^{\nu\lambda}+($quantum corrections$)$. In the
first order in $\hbar$ the equation (\ref{i1}) is equivalent to the Jacobi
identity for the classical matrix  $\omega_{0}^{\mu\nu}$:%
\begin{equation}
\omega_{0}^{\mu\sigma}\partial_{\sigma}\omega_{0}^{\nu\lambda}+\mbox{cycl}(\mu
\nu\lambda)=0,\label{i3}%
\end{equation}
which we have by the construction. In the second order, as well as in all even orders,
the left-hand side of (\ref{i1}) is identically equal to zero, since%

\begin{equation}
f\star_{2n}g-g\star_{2n}f=0.\label{i4}%
\end{equation}
In the third order the condition (\ref{i1}) is not satisfied  for
$\omega^{\mu\nu}=\omega_{0}^{\mu\nu}$, i.e. it does not follow from the Jacobi
identity (\ref{i3}) for $\omega_{0}^{\mu\nu}$. To solve this problem one can
construct a quantum correction to $\omega_{0}$, and this has to be an
$\hbar^{2}$ correction:%
\begin{equation}
\omega^{\mu\nu}=\omega_{0}^{\mu\nu}+\hbar^{2}\omega_{2}^{\mu\nu}+O\left(
\hbar^{4}\right)  \ .\label{i5}%
\end{equation}
Doing so, the third order of the condition (\ref{i1}) will  become%
\begin{equation}
\left(  \xi^{\mu}\star_{3}\omega_{0}^{\nu\lambda}-\omega_{0}^{\nu\lambda}%
\star_{3}\xi^{\mu}\right)  +\left(  \xi^{\mu}\star_{1}\omega_{2}^{\nu\lambda
}-\omega_{2}^{\nu\lambda}\star_{1}\xi^{\mu}\right)  +\mbox{cycl}(\mu\nu
\lambda)=0~.\label{i6}%
\end{equation}
A quantum non-Poisson correction $\omega_{2}^{\mu\nu}$ can be found from
(\ref{i6}) and has the form:%
\begin{equation}
\omega_{2}^{\mu\nu}=\frac{1}{48}\partial_{\gamma}\omega_{0}^{\rho\sigma
}\partial_{\rho}\omega_{0}^{\gamma\delta}\partial_{\sigma}\partial_{\delta
}\omega_{0}^{\mu\nu}-\frac{1}{24}\partial_{\sigma}\partial_{\gamma}\omega
_{0}^{\mu\rho}\partial_{\rho}\partial_{\delta}\omega_{0}^{\nu\sigma}\omega
_{0}^{\gamma\delta}.\label{i7}%
\end{equation}
An explicit formulae for $\omega_{2}^{\mu\nu}$ taking into account the
concrete form (\ref{10}) of $\omega_{0}^{\mu\nu}$\ is presented in appendix.
A systematic procedure for the construction of quantum corrections $\omega_{2n}^{\mu\nu
}$ to the classical Dirac bracket $\omega_{0}^{\mu\nu}$ was described in
\cite{KV}, but explicit calculations were made only up to the fourth order in
$\hbar$ and no general formula is yet available.

Note that in some particular cases in which there is no ordering problem
, e.g., for a linear Poisson structure $\omega^{\mu\nu}$ or if
$\omega^{\mu\nu}$ depends only on one of the coordinates, the quantum Dirac
brackets $\omega^{\mu\nu}$ coincide with the classical ones $\omega_{0}%
^{\mu\nu}$ (there is no corrections). In this case, the Jacobi identity for
the quantum algebra (\ref{11q}) holds true as a consequence of the Jacobi
identity for the matrix $\omega_{0}^{\mu\nu}\left(  x,y\right)  $. The
interesting question is whether it is possible to present an exact formulae
for quantized Dirac brackets of the model or one can only get some reasonable
approximation, expressed as power series in $\hbar$?

To work with operators $\hat{\xi}^{\mu}$ which obey the commutation relations
(\ref{11q}) one can use the polydifferential representation of the algebra
(\ref{11q}): $\hat{\xi}^{\mu}=\xi^{\mu}+i\hbar/2\omega^{\mu\nu}\partial_{\nu
}+...~,$ constructed in \cite{KV}.

\section{Definition of $B_{i}$}

Suppose that we know the position-dependent noncommutativity from some
physical considerations, i.e., the function $d\left(  x,y\right)  ,$ which is
the Weyl symbol of the operator $d\left(  \hat{x},\hat{y}\right)  $, is given.
In order to define the complete algebra (\ref{10}), we need to know the
functions $B_{i}$. For that one can use the equation (\ref{11}). However, one
cannot determine two functions $B_{x}$ and $B_{y}$ from just one equation
(\ref{11}). Therefore, we need to impose one additional condition. We will
consider now two different choices of the additional conditions.

Let us first consider the condition $B_{i}=\varepsilon^{ij}\partial_{j}\phi$,
so that the equation (\ref{11}) becomes%
\[
d=\frac{1}{1+\theta\bigtriangleup\phi}~,
\]
where $\bigtriangleup=\partial_{x}^{2}+\partial_{y}^{2}$. Suppose that the
function $d$ has a rotational symmetry like in the example (\ref{2}), i.e.,%
\begin{equation}
d=\frac{1}{1+\theta f\left(  \alpha\left(  x^{2}+y^{2}\right)  \right)  }~,
\label{11a}%
\end{equation}
where $f$ is some given function, $f\left(  0\right)  =const<\infty$. We will
also need the integral $F$, $F^{\prime}=f,$ $F\left(  0\right)  =const<\infty$.

From (\ref{11}) and (\ref{11a}) one finds
\begin{equation}
\bigtriangleup\phi=f\left(  \alpha\left(  x^{2}+y^{2}\right)  \right)  ~.
\label{12}%
\end{equation}
In polar coordinates $x=r\cos\varphi,\ y=r\sin\varphi$ the equation (\ref{12})
can be written as:%
\begin{equation}
\frac{1}{r}\partial_{r}r\partial_{r}\phi=f\left(  \alpha r^{2}\right)  ~,
\label{13}%
\end{equation}
which yields%
\begin{equation}
\partial_{r}\phi=\frac{F\left(  \alpha r^{2}\right)  }{2\alpha r}+\frac{c}%
{r}~. \label{14}%
\end{equation}
We fix the constant $c$ from the condition%
\begin{equation}
\lim_{\alpha\rightarrow0}\partial_{r}\phi=0 \label{15}%
\end{equation}
which gives $c=-\frac{F\left(  0\right)  }{2\alpha}$
\begin{equation}
\partial_{r}\phi=\frac{F\left(  \alpha r^{2}\right)  -F\left(  0\right)
}{2\alpha r}. \label{16}%
\end{equation}
Then we calculate%
\begin{align}
&  B_{x}=\partial_{y}\phi=\left(  \sin\varphi\partial_{r}+\frac{1}{r}%
\cos\varphi\partial_{\varphi}\right)  \phi\left(  r\right)  =\label{17}\\
&  \sin\varphi\frac{F\left(  \alpha r^{2}\right)  -F\left(  0\right)
}{2\alpha r}~=y\frac{F\left(  \alpha\left(  x^{2}+y^{2}\right)  \right)
-F\left(  0\right)  }{2\alpha\left(  x^{2}+y^{2}\right)  },\nonumber
\end{align}
and%
\begin{align}
&  B_{y}=-\partial_{x}\phi=-\left(  \cos\varphi\partial_{r}-\frac{1}{r}%
\sin\varphi\partial_{\varphi}\right)  \phi\left(  r\right)  =\label{18}\\
&  -\cos\varphi\frac{F\left(  \alpha r^{2}\right)  -F\left(  0\right)
}{2\alpha r}~=-x\frac{F\left(  \alpha\left(  x^{2}+y^{2}\right)  \right)
-F\left(  0\right)  }{2\alpha\left(  x^{2}+y^{2}\right)  }.\nonumber
\end{align}
We see that $B_{i}\rightarrow0$ when $\alpha\rightarrow0$.

The second choice is $B_{x}=B_{y}=\chi$. Note, that this condition implies
that $\left\{  p_{x},p_{y}\right\}  _{D}=0$. We consider more general case%
\[
d=\frac{1}{1+\theta g\left(  \alpha,x,y\right)  },
\]
where $g\left(  \alpha,x,y\right)  $ is an arbitrary function, $g\left(
0,x,y\right)  =0$. The equation (\ref{11}) yields%
\[
\left(  \partial_{x}-\partial_{y}\right)  \chi=g\left(  \alpha,x,y\right)
\]
After the change of variables $\xi=x-y,\ \eta=x+y,$ one has%
\[
\partial_{\xi}\chi=g\left(  \alpha,\frac{1}{2}\left(  \xi+\eta\right)
,\frac{1}{2}\left(  \xi-\eta\right)  \right)  ,
\]
the solution of this equation is%
\[
\chi=G_{\xi}\left(  \xi,\eta\right)  +G_{0}\left(  \eta\right)
\]
where%
\[
G_{\xi}\left(  \xi,\eta\right)  =%
{\displaystyle\int}
d\xi g\left(  \alpha,\frac{1}{2}\left(  \xi+\eta\right)  ,\frac{1}{2}\left(
\xi-\eta\right)  \right)  ,
\]
and the function $G_{0}\left(  \eta\right)  $ can be determined from the
condition that $\lim_{\alpha\rightarrow0}\chi=0$.

Thus, we have constructed the classical model (\ref{8}) which after
quantization leads to the two-dimensional QM with position-dependent
noncommutativity $[\hat{x},\hat{y}]=i\theta d\left(  \hat{x},\hat{y}\right)
$. To define this model we use the position-dependent noncommutativity itself,
which is supposed to be known ab initio, and an additional condition, imposed
by hand from some physical considerations. For example, if we want $[\hat
{p}_{x},\hat{p}_{y}]=0$, we choose the additional condition $B_{x}=B_{y}$, etc.

\section{Local noncommutativity}

Let us consider the particular example of local noncommutativity (\ref{2}). In
this case the function $d$ is%
\[
d=\frac{1}{1+\theta\alpha\left(  x^{2}+y^{2}\right)  }.
\]
The first choice of additional condition ($B_{i}=\varepsilon^{ij}\partial
_{j}\phi$) implies:%
\[
B_{x}=-\frac{\alpha}{4}y\left(  x^{2}+y^{2}\right)  ,\ \ B_{y}=\frac{\alpha
}{4}x\left(  x^{2}+y^{2}\right)  ,
\]
and the Dirac brackets (\ref{10}) are%
\begin{align}
&  \left\{  x,y\right\}  _{D}=\theta d~,\ \ \left\{  p_{x},p_{y}\right\}
_{D}=\frac{3\theta\alpha^{2}}{16}\left(  x^{2}+y^{2}\right)  ^{2}%
d~,\label{19}\\
&  \left\{  x,p_{x}\right\}  _{D}=\left[  1+\frac{\alpha\theta}{4}\left(
x^{2}+3y^{2}\right)  \right]  d~,\ \ \left\{  x,p_{y}\right\}  _{D}%
=-\frac{\alpha\theta}{2}xyd~,\nonumber\\
&  \left\{  y,p_{y}\right\}  _{D}=\left[  1+\frac{\alpha\theta}{4}\left(
3x^{2}+y^{2}\right)  \right]  d~,\ \ \left\{  y,p_{x}\right\}  _{D}%
=-\frac{\alpha\theta}{2}xyd.\nonumber
\end{align}
The second choice means%
\[
B_{x}=B_{y}=\frac{\alpha}{3}\left(  x^{3}-y^{3}\right)  ~,
\]
and%
\begin{align}
&  \left\{  x,y\right\}  _{D}=\theta d~,\ \ \left\{  p_{x},p_{y}\right\}
_{D}=0,\label{19a}\\
&  \left\{  x,p_{x}\right\}  _{D}=\left[  1+\alpha\theta y^{2}\right]
d~,\ \ \left\{  x,p_{y}\right\}  _{D}=\alpha\theta x^{2}d~,\nonumber\\
&  \left\{  y,p_{y}\right\}  _{D}=\left[  1+\alpha\theta x^{2}\right]
d~,\ \ \left\{  y,p_{x}\right\}  _{D}=\alpha\theta y^{2}d.\nonumber
\end{align}

In order to compare the two models we consider the limit $r\rightarrow\infty$.
In both cases $\left\{  x,y\right\}  _{D}\rightarrow0$ and the Dirac brackets
$\left\{  x,p_{x}\right\}  _{D},\ \left\{  x,p_{y}\right\}  _{D},\ \left\{
y,p_{y}\right\}  _{D}$ and $\left\{  y,p_{x}\right\}  _{D}$ are limited
functions in this limit. However, $\lim_{r\rightarrow\infty}\left\{
p_{x},p_{y}\right\}  _{D}=\infty$ in the first model, while $\left\{
p_{x},p_{y}\right\}  _{D}=0$ in the second. Since, usually, the non-zero
commutator of the momenta means the presence of a magnetic field, it would be
difficult to give some physical meaning to the first model on the infinity
whereas the second one is free from this difficulty.

\section{Discussions and conclusions}

We have proposed a model of the consistent quantum mechanics with
position-dependent noncommutativity. Our construction is based on the
first-order Lagrangian, which after quantization reproduces the desired
commutation relations between the operators of coordinates and momenta.

Note that a first-order Lagrangian for the Duval-Horvathy model \cite{Duval}
can also lead to the position-dependent Dirac brackets \cite{Lukierski1}, see
also \cite{Horvathy}, where the correspondent symplectic structure was
obtained by means of introducing an interaction with the magnetic field in the
model of nonrelativistic anyon \cite{Horvathy1}. However, the
position-dependence in this case is due to the presence of a nonconstant
magnetic field $B\left(  x\right)  $. In our model (\ref{8}) the
noncommutativity is caused by other factors and magnetic field can enter the
theory via Hamiltonian $H\left(  x,p\right)  $. Also, the possibility to
localize the noncommutativity within the model \cite{Lukierski} meets some
difficulties, since the magnetic field $B\left(  x\right)  $ should go to
infinity outside the area of local noncommutativity. Three-dimensional
generalization of the model \cite{Lukierski} was considered in
\cite{Chaichian}.

It should be mentioned that the particular case of a position-dependent
noncommutativity, a model of a point particle on kappa-Minkowski space was
derived from a first-order Lagrangian in \cite{Ghosh}.

In order to obtain some phenomenological consequences of such a type of
noncommutativity in space it would be interesting to consider some particular
physical problems in the presence of this noncommutativity. For example, the
scattering of plane waves on the local noncommutativity. For that one needs to
take the Hamiltonian of free particle $\hat{H}=\frac{1}{2}\left(  \hat{p}%
_{x}^{2}+\hat{p}_{y}^{2}\right)  $ and to use perturbation theory on $\theta$.
Also, it would be interesting to calculate the uncertainty relations.

\section*{Acknowledgements}

We are grateful to Dmitri Vassilevich for fruitful discussions. We also thank
Funda\c{c}\~{a}o de Amparo \`{a} Pesquisa do Estado de S\~{a}o Paulo (FAPESP)
and Conselho Nacional de Desenvolvimento Cient\'{\i}fico e Tecnol\'{o}gico
(CNPq) for partial support.

\section{Appendix}

Taking into account the concrete form (\ref{10}) of $\omega_{0}^{\mu\nu}$ one
can calculate the explicit form of quantum non-Poisson correction $\omega
_{2}^{\mu\nu}$, which are listed below with $\mu<\nu$:%

\begin{align*}
\omega_{2}^{12} &  =\frac{\theta^{3}}{24}\left[  \frac{1}{2}\left(
\partial_{2}d\right)  ^{2}\partial_{1}^{2}d-\partial_{1}d\partial_{2}%
d\partial_{1}\partial_{2}d+\frac{1}{2}\left(  \partial_{1}d\right)
^{2}\partial_{2}^{2}d\right.  \\
&  +\left.  d\left(  \partial_{1}\partial_{2}d\right)  ^{2}-d\partial_{2}%
^{2}d\partial_{1}^{2}d\right]  ,\\
\omega_{2}^{ij+2} &  =\frac{\theta^{2}}{24}\left[  \frac{1}{2}\left(
\partial_{2}d\right)  ^{2}\partial_{1}^{2}-\partial_{1}d\partial_{2}%
d\partial_{1}\partial_{2}+\frac{1}{2}\left(  \partial_{1}d\right)
^{2}\partial_{2}^{2}\right]  \\
&  \times\left(  \delta_{j}^{i}d-\theta\varepsilon^{ik}\partial_{k}%
B_{j}d\right)  -\frac{\theta^{2}}{24}\varepsilon^{im}d\left(  \partial
_{j}\partial_{1}d\partial_{m}\partial_{2}d-\partial_{j}\partial_{2}%
d\partial_{m}\partial_{1}d\right)  \\
&  +\frac{\theta^{3}}{24}\varepsilon^{im}\varepsilon^{jk}d\left[  \partial
_{n}\partial_{1}d\partial_{m}\partial_{2}\left(  \partial_{k}B_{n}d\right)
-\partial_{n}\partial_{2}d\partial_{m}\partial_{1}\left(  \partial_{k}%
B_{n}d\right)  \right]  ,\\
\omega_{2}^{34} &  =\frac{\theta^{3}}{24}\left[  \frac{1}{2}\left(
\partial_{2}d\right)  ^{2}\partial_{1}^{2}-\partial_{1}d\partial_{2}%
d\partial_{1}\partial_{2}+\frac{1}{2}\left(  \partial_{1}d\right)
^{2}\partial_{2}^{2}\right]  \\
&  \times\left(  \left(  \partial_{2}B_{2}\partial_{1}B_{1}-\partial_{1}%
B_{2}\partial_{2}B_{1}\right)  d\right)  -\\
&  \frac{\theta}{24}d\left[  \partial_{n}\partial_{1}\left(  \delta_{m}%
^{1}d-\theta\partial_{2}B_{m}d\right)  \partial_{m}\partial_{2}\left(
\delta_{n}^{2}d+\theta\partial_{1}B_{n}d\right)  \right.  \\
&  -\left.  \partial_{n}\partial_{2}\left(  \delta_{m}^{1}d-\theta\partial
_{2}B_{m}d\right)  \partial_{m}\partial_{1}\left(  \delta_{n}^{2}%
d-\theta\partial_{1}B_{n}d\right)  \right]  .
\end{align*}

\end{document}